\documentclass{article}[12pt]

\usepackage{amsmath,amssymb} \usepackage{graphicx}
\def\eq{\begin{equation}}
\def\en{\end{equation}}
\def\eqa{\begin{eqnarray}}
\def\ena{\end{eqnarray}}
\newcommand{\me}{\mathrm{e}}
\newcommand{\dif}{\mathrm{d}}

\def\expval#1{\langle \, #1 \,\rangle}
\def\expvalc#1{\expval{#1}_{c}}
\def\expvalequil#1{\expval{#1}_{\mathit{eq}}}

\def\anticomm#1#2{\{#1 ,\, #2  \}}

\def\Qhat{\hat{Q}}

\def\phiaB{\phi_{a}}
\def\phiaF{\hat\phi_{a}}
\def\phibF{\hat\phi_{b}}
\def\gsusy{g^{{\scriptstyle S}}}
\def\bulk{^{\rm bulk}}
\def\tot{}
\def\thefootnote{\fnsymbol{footnote}} 
\begin{document} 
\begin{titlepage}
%\noindent October 2, 2008          \hfill\\
%bsusyDF1002.tex
\begin{center}
%\hfill hep-th/yymmnnn  \\

\vskip .5in \renewcommand{\thefootnote}{\fnsymbol{footnote}}
{\Large \bf General properties of the boundary 
renormalization group flow\\[1.5ex] 
for supersymmetric systems in 1+1 dimensions}

%\vskip .50in

\vskip .5in {\large Daniel Friedan}${}^{1,2,}$\footnote{email address: friedan@physics.rutgers.edu} and
{\large Anatoly Konechny}${}^{3,4,}$\footnote{email address: anatolyk@ma.hw.ac.uk}

\vskip 0.5cm
{\large \em ${}^{1}$Department of Physics and Astronomy,\\
Rutgers, The State University of New Jersey,\\
Piscataway, New Jersey 08854-8019 U.S.A. \\

${}^{2}$Natural Science Institute, 
The University of Iceland, 
Reykjavik, Iceland\\

${}^{3}$Department of Mathematics,
Heriot-Watt University,\\
Riccarton, Edinburgh, EH14 4AS, UK\\

${}^{4}$Maxwell Institute for Mathematical Sciences, 
Edinburgh, UK}\\
\end{center}

\vskip .5in

\begin{abstract} \large
%insert abstract here
We consider the general supersymmetric one-dimensional quantum system 
with boundary,
critical in the bulk but not at the boundary.
The renormalization group flow on the space of 
boundary conditions is generated by the boundary beta functions 
$\beta^{a}(\lambda)$ for the boundary coupling constants $\lambda^{a}$.
We prove a gradient formula
$\partial\ln z/\partial\lambda^{a} =-\gsusy_{ab}\beta^{b}$ 
where $z(\lambda)$ is the boundary partition function
at given temperature $T=1/\beta$,
and $\gsusy_{ab}(\lambda)$ is a certain positive-definite metric on the space of 
supersymmetric boundary conditions.
The proof depends on canonical ultraviolet behavior at the boundary.
Any system whose short distance behavior is governed by a fixed point 
satisfies this requirement.
The gradient formula implies
that the boundary energy,
$-\partial\ln z/\partial\beta = -T\beta^{a}\partial_{a}\ln z$,
is nonnegative.
Equivalently,
the quantity $\ln z(\lambda)$ decreases under the renormalization 
group flow.

\end{abstract}
\end{titlepage}
\large
\newpage
\renewcommand{\thepage}{\arabic{page}} \setcounter{page}{1} \setcounter{footnote}{0}
\renewcommand{\thefootnote}{\arabic{footnote}}
%%%%%%%%%%%

\large 
\section{Introduction}
\renewcommand{\theequation}{\arabic{section}.\arabic{equation}}
\setcounter{equation}{0}
In this paper we consider the renormalization group flow
for supersymmetric one-dimensional quantum systems with
boundary which are critical in the bulk but not critical on the
boundary.
First
we give a brief overview of what is known without the assumption of supersymmetry. 
%NEW
There are many condensed matter applications for such systems,
such as quantum impurities and quantum Hall edge excitations
(see e.g.\ \cite{Saleur} for a review).  
%%%%END OF NEW
%NEW-2008.09.22
We expect that supersymmetric bulk-critical one-dimensional systems with boundaries 
-- and junctions -- can be realized in practice.
Such supersymmetric quantum circuits
might be useful for large-scale quantum computing \cite{Friedan:2005bz}.
%%%%END OF NEW-2008.09.22

Consider a bounded system of length $L$ at 
low temperature $T=1/\beta$.
Let $H_{L}$ be the 
hamiltonian\footnote{We are considering 1d quantum mechanical systems
so we can assume unitarity:
the hamiltonian is a self-adjoint operator acting on a 
Hilbert space of states.  By Wick rotation, our results apply 
equally well to 2d statistical systems that satisfy reflection positivity.} of the bounded system.
The partition function is
$ Z_{L}= {\rm tr}\left ( e^{-\beta H_{L}}\right )$.
There are two boundaries, one at each end.
In the limit $L\to \infty$, the two boundaries 
decouple and the partition function of the whole system factorizes into
a bulk contribution and two boundary contributions:
\begin{equation}
Z_{L} \sim e^{\pi c L/6\beta}zz' \, .
\end{equation}
Here $c$ is the central charge of the conformal field theory describing the bulk critical 
system, $-\pi c/6\beta^{2}$ is the universal free energy density of the 
bulk conformal field theory,
and  $z$ and $z'$ are the $L$-independent contributions of the boundaries. For  
a unitary theory the sign of $z$ can be fixed so that $z$ is positive.
% At non-zero temperature supersymmetry in such systems is spontaneously broken. 
%The logarithm of  the partition function for 
 %$L\gg \beta=1/T$ can be written as 
%\begin{equation}
% \ln Z = \ln z + \frac{c\pi L}{6\beta} + {\cal O}(e^{-L/\beta})\, .
%\end{equation}
%, $\ln z$ is the $L$-independent piece. 
The quantity $z$ is the boundary partition function.
%NEW-2008.09.22
It is a function
$
z(\lambda,\mu\beta)
$
depending on the boundary coupling constants $\lambda^{a}$
that parametrize the boundary condition
and on the temperature $T=1/\beta$
(in dimensionless units of the energy scale $\mu$).
%END OF NEW-2008.09.22

The boundary partition function has no representation of the form
$z={\rm tr}\left ( e^{-\beta h}\right )$
so there is no reason to believe that the boundary thermodynamic 
functions constructed from $z$
will satisfy the usual thermodynamic principles.
Nevertheless, it can be proved \cite{FK} that the boundary entropy 
\begin{equation}\label{s}
s=(1-\beta\frac{\partial}{\partial \beta})\ln z
\end{equation}
does decrease monotonically with temperature.
That is, the boundary satisfies the second law of thermodynamics. 
We emphasize that this was not necessarily to be expected.
The entropy of the whole system behaves as
\begin{equation}
S_{L} \sim s + s' + \frac{c\pi L}{3\beta}
\end{equation}
as $L\rightarrow \infty$.
The total entropy $S_{L}$ decreases monotonically with temperature, 
but so does the bulk term.
The subtraction of the bulk term
precludes a straightforward derivation of the second law of
thermodynamics for the boundary entropy $s$.

The renormalization 
group (RG) equation is
\eq
\mu \frac{\partial \ln z}{\partial\mu}
= \beta^{a}\frac{\partial \ln z}{\partial\lambda^{a}}
\en
where the $\beta^{a}(\lambda)$ are the boundary beta functions.
The critical boundary conditions are described by the fixed points,
$\beta^{a}=0$.
The boundary partition function at a fixed point is a number, 
scale invariant and therefore independent of temperature,
traditionally denoted $z=g$.
The number $g$ was introduced as an invariant of critical boundary systems
by Affleck and Ludwig \cite{AL1},
who called it the universal noninteger ground state degeneracy.
They conjectured 
\cite{AL1,AL2} that, for two critical boundary conditions  
connected by an RG trajectory, 
the value of
$g$ at the infrared fixed point is
always smaller than the value 
at the ultraviolet fixed point. 
Affleck and Ludwig's conjecture follows from
the second law of boundary thermodynamics,
because $s=\ln g$ at each of the fixed points,
and the scale $\mu$ can be traded for the temperature.

The second law of boundary thermodynamics
is a consequence of yet a stronger statement, the 
boundary gradient formula proved in \cite{FK}:
\begin{equation}\label{bosonicgf}
\frac{\partial s}{\partial  \lambda^{a}} = -g_{ab}\beta^{b}
\end{equation}
where $g_{ab}$ is a certain positive definite metric on the space
of boundary couplings.
Since $\ln z$ and $s$ depend on the dimensionless product $\mu\beta$,
the RG equation for $s$ can be written
\eq
\mu \frac{\partial s}{\partial\mu} =\beta\frac{\partial s}{\partial\beta}
= \beta^{a}\frac{\partial s}{\partial\lambda^{a}} \, .
\en
Contracting (\ref{bosonicgf}) with $\beta^{a}$ gives
\begin{equation}
\beta\frac{\partial s}{\partial\beta} = 
\beta^{a}\frac{\partial s}{\partial\lambda^{a}}= -\beta^{a}g_{ab}\beta^{b}\le 0
\end{equation}
which says that $s$ decreases as the temperature decreases.
The boundary second law thus follows from
the gradient formula.

%NEW-2008.09.22
The proof of the gradient formula given in \cite{FK}
used the euclidean description of the
finite temperature quantum system.
The metric in equation (\ref{bosonicgf}) is
\begin{equation}\label{bosonicmetric}
g_{ab}=\beta \int_{0}^{\beta} d\tau  \;
\left [ 1-\cos\left(2\pi\tau/\beta\right)\right ]
\,
\langle\phi_{a}(\tau)\phi_{b}(0)\rangle_{c}\, 
\end{equation}  
where  $\expvalc{\cdots}$ stand for  the connected thermal correlation functions.
The one-dimensional system with a single boundary
is described by a two-dimensional euclidean field theory
with spatial coordinate $x$, $0\le x< \infty$,
and euclidean time $\tau$.
The boundary is at $x=0$.
The euclidean time $\tau$ is
periodic with period $\beta$.
The euclidean space-time is the semi-infinite cylinder
with coordinates $(x,\tau)$.
The boundary coupling constants $\lambda^{a}$ couple to boundary 
operators $\phi_{a}(\tau)$, localized at $x=0$, so that
\begin{equation}\label{ccs}
\frac{\partial \ln z}{\partial \lambda^{a}} = 
\int_{0}^{\beta} d\tau \; \langle \phi_{a}(\tau)\rangle 
=\beta \langle \phi_{a}\rangle \, .
\end{equation}
An alternative proof of the gradient formula (\ref{bosonicgf}) using
real time methods was presented in \cite{FK2}.  There, the metric $g_{ab}$
was expressed via response functions.
%NEW-2008.09.22 END
The proof of the gradient formula (\ref{bosonicgf}) relies on the
assumption that the two-point correlation functions
of the boundary operators $\phi_{a}(\tau)$
with themselves and with the stress-energy tensor
and $T_{\mu\nu}(x,\tau)$
behave canonically at short distance.
This assumption is valid
if the ultraviolet limit is governed by a fixed point,
because then the boundary operators $\phi_{a}(\tau)$ must be relevant 
at the fixed point.
It is interesting to note that no
assumption of this kind is needed to prove
Zamolodchikov's $c$-theorem \cite{Zam},
which establishes
the monotonic decrease of the $c$-function under the RG flow in
the space of bulk 2d field theories.

Now we specialize to supersymmetric one dimensional systems with 
boundary.
%NEW-2008.09.22
In supersymmetric systems the thermodynamic energy 
$-\partial\ln Z/\partial \beta$ is always nonnegative,
because the hamiltonian is of the form $H=\Qhat^{2}$,
where $\Qhat$ is the supercharge operator.
However it is not obvious that the
boundary energy in such a supersymmetric system
should be nonnegative.
%NEW-2008.09.22 NEW
Consider again a finite system of length $L$.
For the whole system, certainly $-\partial\ln Z_{L}/\partial\beta \ge 0$,
but
\begin{equation}
-\frac{\partial \ln Z_{L}}{\partial \beta}=-\frac{\partial \ln z}{\partial \beta}
-\frac{\partial \ln z'}{\partial \beta}
+\frac{\pi cL}{6\beta^{2}}\, 
\end{equation}
as $L\to \infty$.  The positivity of the large bulk energy prevents
us from concluding that the boundary energy is positive.

In this paper we prove the positivity of the boundary energy by 
deriving a new gradient formula for the supersymmetric boundary RG flow
\begin{equation}\label{susygradf}
\frac{\partial \ln z}{\partial  \lambda^{a}} = -\gsusy_{ab}\beta^{b}
\end{equation}
where $\gsusy_{ab}$ is a certain positive-definite metric on the space of supersymmetric
boundary conditions (not the same metric as in the general gradient 
formula).
Contracting with $\beta^{a}$ gives
\begin{equation}
-\frac{\partial \ln z}{\partial \beta}=T\beta^{a}\gsusy_{ab}\beta^{b}\ge 0
\end{equation}
which proves that the boundary energy is nonnegative.

As in the case of the general gradient formula (\ref{bosonicgf}), which was to a large extent 
inspired by work done in string theory \cite{Witten1, Witten2, 
Shat1, Shat2, KMM1}, 
the existence of a different gradient formula for the supersymmetric 
boundary RG flow
was anticipated in the string theory literature 
\cite{KMM2,M,NP}. 
It was conjectured in \cite{KMM2,M,NP} that $z$ is a potential function for 
such a gradient formula\footnote{In string theory, one wants
a gradient formula for the beta-function, such as (\ref{susygradf}),
in order to have a space-time action principle.
In string theory it is $z$ rather than $\ln z$ 
that is a natural potential function (a string field theory action).
The link between (\ref{bosonicgf}) and its stringy version
requires special treatment of the tachyon zero mode \cite{FK}.
The stringy version of the supersymmetric gradient formula 
(\ref{susygradf}) is trivially obtained by 
multiplying both sides by $z$.}.   In \cite{NP} the expression 
for the metric $g_{ab}^{\rm S}$ was put forward, 
%NEW-2008.09.22
which we will show to be correct, but a proof of the gradient formula was still lacking.
%NEW-2008.09.22 END
In this paper we give two different proofs of (\ref{susygradf}). 
In section \ref{sect:euclidean_proof} we give a proof using the formalism of euclidean quantum field theory.
In section \ref{sect:realtime_proof} we use real time methods. The two proofs are compared 
in section 5. In the euclidean approach the 
metric is written
\begin{equation}\label{susymetric}
\gsusy_{ab}=2\pi \int_{0}^{\beta} d\tau  \;
\sin\left( \pi\tau/\beta\right)
\langle \hat \phi_{a}(\tau)\hat \phi_{b}(0)\rangle 
\end{equation}
where the $\hat \phi_{a}(\tau)$ are the fermionic superpartners\footnote{The one-point 
functions $\langle \phi_{a}(\tau)\rangle$ which appear on the left hand 
side of the gradient formula can be non-vanishing 
because the global supersymmetry is spontaneously broken at non-zero temperature.} of the 
bosonic boundary operators $\phi_{a}(\tau)$. 
In the real time approach,
the same metric is written in terms of real time response functions
of the $\phiaF(t)$,\footnote{We abuse notation in writing $\hat\phi_{b}(\tau)$ when we are 
discussing physics in euclidean time,
and $\hat\phi_{b}(t)$ when discussing real time physics.
To be consistent, we should write either $\hat\phi_{b}(\tau)$ and 
$\hat\phi_{b}(i t)$ or $\hat\phi_{b}(-it)$ and 
$\hat\phi_{b}(t)$.
We are perhaps also abusing terminology when we refer to \emph{response functions}
of fermionic operators.}
\eq
\label{susymetricrealtime}
\gsusy_{ab}=\pi \int_{-\infty}^{\infty} dt\;
e^{-\pi |t|/\beta}\expval{
\anticomm{\hat\phi_{b}(t)}{\hat\phi_{a}(0)}}
\,.
\en

Like the general gradient formula,
%NEW-2008.09.22
formula (\ref{susygradf}) is proved under the condition
of canonical short distance behavior at the boundary,
now for the correlation functions
$\langle \hat \phi_{a}(\tau)\hat \phi_{b}(\tau')\rangle$,
$\langle \hat \phi_{a}(\tau)\hat \theta(\tau')\rangle$,
and 
$\langle G_{\mu r}(\tau, x)\hat \phi_{b}(\tau')\rangle$
where $ G_{\mu r}$ is the bulk supersymmetry current
and $\hat \theta$ is its boundary part.
Again, the condition is satisfied
if the extreme UV limit is 
described by a fixed point (which would necessarily be supersymmetric).
Then the UV scaling dimension 
of $\hat \phi_{a}$ is at most $1/2$ and the bulk supercurrent 
$G_{\mu r}$ has canonical scaling dimension $3/2$. 
%NEW-2008.09.22 END
At present, we see only technical reasons
for the gradient formulas to depend on canonical UV behavior
at the boundary.

The metric $\gsusy_{ab}(\lambda)$, like the bosonic metric 
$g_{ab}(\lambda)$,
is covariant under change of coordinates $\lambda^{a}$ in the 
space of boundary conditions.
This follows from formulas (\ref{bosonicmetric}) and (\ref{susymetric}) 
where the metrics are defined by expressions which are
insensitive to possible contact terms in the two 
point functions.

However both metrics may fail to be invariant under the RG flow.
RG invariance is the condition that
change of scale is equivalent to flow under the RG,
\eq
\mu \frac{\partial g_{ab}}{\partial \mu} 
= (\mathcal{L}_{\beta} g)_{ab} = 
\beta^{c}\frac{\partial g_{ab}}{\partial \lambda^{c}} +
\frac{\partial \beta^{c}}{\partial \lambda^{a}}g_{cb} + 
g_{ac}\frac{\partial \beta^{c}}{\partial \lambda^{b}}
\,.
\en
RG invariance means that the metric, though it is defined at a 
certain temperature (scale), in fact does not depend on the arbitrary choice 
of scale.  The metric depends only on the running coupling constants
at the temperature at which it is measured.
Without RG invariance, the metric depends on more than the running 
coupling constants at the physical temperature.  There are many 
different gradient formulas, one for each temperature, all satisfied.
We suppose that this unsatisfactory situation might be alleviated by 
introduction of some auxiliary couplings.

The problem with RG invariance of the metric is that the local fields 
need only transform covariantly under the RG flow up to total 
derivative operators,
\begin{equation} \label{RGop}
\mu \frac{\partial  \phi_{a}(\tau)}{\partial \mu} = 
\frac{\partial\beta^{b}}{\partial\lambda^{a}}\phi_{b}(\tau) + 
\partial_{\tau}\chi_{a}(\tau)\, . 
\end{equation}
Such admixtures
do not affect such quantities as $\partial s/\partial\lambda^{a}$
and $\partial \ln z/\partial\lambda^{a}$
but do affect
local correlators such as are used 
in the definition of the metric (\ref{bosonicmetric}).
The transformation law (\ref{RGop}) is consistent with our UV assumptions
as long as the UV scaling dimension of the field $\chi_{a}$ is zero.
Such fields can exist if the UV fixed point theory
has multiple -- degenerate -- ground states.
This is in the ultraviolet limit, not in the infrared,
so there is no physical pathology.
Note that the left hand side of the gradient formula is RG invariant,
so the right hand side, $g_{ab}\beta^{b}$, must also be RG invariant.
This puts constraints on  the correlators of the $\chi_{a}(\tau)$.
For supersymmetric theories, the scale transformation of the metric $g_{ab}^{\rm S}$
is affected by analogous admixtures in the RG transformation law for 
the fermionic boundary fields,
\begin{equation} \label{RGop2}
\mu \frac{\partial  \hat \phi_{a}(\tau)}{\partial \mu} = \frac{\partial\beta^{b}}{\partial\lambda^{a}}
\hat \phi_{b}(\tau) + \{ \hat Q, \chi_{a}(\tau)\} \,.
\end{equation}
It would be desirable both to find explicit examples where the metric is 
not RG invariant and also to get a deeper general understanding of such situations.

In an isolated supersymmetric system,
the ground state energy $E_{0}$ is zero
if and only if
the supersymmetry is unbroken in the ground state.
The low temperature limit
of the partition function
is therefore a definitive diagnostic of spontaneous supersymmetry breaking in 
the ground state.
When the supersymmetry is broken, then $\ln Z$ decreases
as $-\beta E_{0}$, with no lower bound.
When the supersymmetry is unbroken,
the partition function $Z$ decreases to a lower bound,
the ground state degeneracy, so $\ln Z \ge 0$.
In supersymmetric boundary systems, the low temperature limit of $\ln
z$ is more problematic.
The gradient formula we prove here, equation (\ref{susygradf}),
implies that the boundary thermodynamic energy
is nonnegative
\eq
e(\beta) = -\frac{\partial \ln z}{\partial \beta} \ge 0
\,.
\en
The general gradient formula implies the second law for the boundary,
\eq
\frac{\partial e}{\partial \beta}
= - \frac{\partial^{2}\ln z}{\partial \beta^{2}}
= \frac1\beta \frac{\partial s}{\partial \beta}
\le 0
\,.
\en
%NEW-2008.09.22
So the
boundary energy is nonnegative and decreases monotonically as $\beta\rightarrow \infty$.
%NEW-2008.09.22 END
Therefore it must have a nonnegative limit
\eq
\lim_{\beta\rightarrow\infty} e(\beta) = e_{0} \ge 0
\,.
\en
The bulk superconformal invariance implies that there is no bulk ground state 
energy,
so all the ground state energy must be localized in the boundary.
%NEW-2008.09.22
The total ground state energy is $e_{0}$.
Therefore the supersymmetry is spontaneously broken if and only if $e_{0}>0$.
Certainly, if $e_{0}>0$ then $\ln z$ goes as $-\beta 
e_{0}$ for large $\beta$.
%NEW-2008.09.22
When the supersymmetry is unbroken, $e_{0}=0$,
we can ask if $\ln z \ge 0$ as $\beta\rightarrow\infty$,
as for an isolated supersymmetric system.
The elementary proof does not work, as before,
because in the finite system
\eq
\ln Z_{L} \sim \ln z + \ln z' + \frac{\pi c L}{6 \beta}
\en
so $\ln z$ is the difference of two positive numbers.\footnote{Note
that the limits $L\rightarrow \infty$ and 
$\beta\rightarrow\infty$ do not commute.}
In fact, an example of 
supersymmetric critical boundary with $\ln z<0$ has been given in 
\cite{Nep} (the boundary condition labeled `0' there).

%NEW-2008.09.22 END
We cannot even say whether or not $\ln z$ is bounded below 
as $\beta\rightarrow \infty$, in general.
There seems to be a parallel with the question of a lower bound on 
the boundary entropy $s$ in the general, non-supersymmetric case.
Unlike ordinary entropy, $s$ can be negative.
There are many examples.
We cannot prove a universal lower bound on $s$, or a lower bound
for a given bulk conformal field theory.
We cannot even prove that $s$ is bounded below as a function of 
$\beta$ for a given boundary system.
Some partial results were found in \cite{FK2}.
It does not seem that supersymmetry helps to get any stronger results
on a lower bound for $s$.
The methods of \cite{FK2} can be
easily generalized to study the rate of change of the boundary free energy at
low temperature in the supersymmetric case, but again we find nothing conclusive.
New methods are needed to put a definite lower
bound either on $s$ or on $\ln z$.
The second law of boundary thermodynamics, which holds in 
general,
and the positivity of the
boundary energy for supersymmetric systems
both suggest that boundaries of systems critical in the bulk behave 
in some respects like isolated thermodynamic systems.
The absence of lower bounds on $s$ and $\ln z$ 
would weaken this analogy.
The absence of lower bounds also prevents the gradient formula from 
definitively controlling the infrared limits of the boundary renormalization group.
%NEW-2008.09.22 END

Finally,
it would be desirable to have some
physical insight into the crucial roles of bulk conformal invariance
and canonical UV boundary behavior
in the picture of boundary physics that is
provided by the two gradient formulas.

\section{Supersymmetry in the presence of a boundary in 2d and 1+1d}
\renewcommand{\theequation}{\arabic{section}.\arabic{equation}}
\setcounter{equation}{0}
%%%%%%%%%5555555%%%%
%%%%%%%%%%%%%%%%%%%%%%%%%%%%%%%%%%%%%%%%%%%%%%%%%%%%% SPINORS %%%%%%%%%%%%%%%%%%%%%%%%%%%%%%%%%%%
A near critical one-dimensional quantum system with boundary, at 
temperature $T=1/\beta$, can be described
by a two-dimensional Euclidean quantum field theory on a semi-infinite cylinder 
with coordinates $(x,\tau)$, as defined in the introduction.
%The coordinates on a cylinder are $\tau \sim \tau + \beta$, $x\ge 0$ where $\beta=1/T$ %is the
%inverse temperature.
Space is the half line $0\le x < \infty$.
Correlation functions of bosonic fields are periodic in euclidean 
time $\tau$, with period $\beta$, while correlation functions of fermionic
fields are anti-periodic.
The Wick rotation to real time is given by $\tau = it$.\footnote{Again, we will abuse
notation by writing fields and operators as functions of $\tau$ 
working in euclidean time,
and as functions of $t$ when working in real time.}
It is convenient to introduce a complex coordinate 
$w=x+i\tau=x-t$,
and its complex conjugate
$\bar w =x-i\tau=x+t$.
We set the RG scale $\mu$ to $1$,
since variation of the RG scale
is equivalent to variation of $\beta$.

\subsection{Spinor conventions}
A Dirac spinor $\hat\epsilon$ in two dimensions has two complex components
\begin{equation}
\hat\epsilon= \left(\begin{array}{c}
\hat\epsilon_{+}\\ \hat\epsilon_{-}
\end{array}\right)
\end{equation}
where $\hat\epsilon_{+}$ and $\hat\epsilon_{-}$ are the positive and negative 
chirality components.
The euclidean reality condition is $(\hat\epsilon_{+})^{*} = 
\hat\epsilon_{-}$.
We use
$\mu, \nu,\ldots$ for vector indices and
$r,s, \dots$ for spinor indices.
Spinor indices are raised
and lowered according to the rule $\hat\epsilon^{+}=2\hat\epsilon_{-}$,
$\hat\epsilon^{-}=2\hat\epsilon_{+}$.
Our Dirac matrices $\gamma^{\mu}$ are
\eq
\gamma^{w}=\gamma^{x}-\gamma^{t}= \left( \begin{array}{cc}
0& 2i\\
0&0
\end{array} \right) \,,
\quad (\gamma^{w})^{-}_{+} = 2i,\quad
\gamma^{w}_{++} = i\,,
\en
\eq
\gamma^{\bar w}=\gamma^{x}+\gamma^{t}=\left( \begin{array}{cc}
0& 0\\
-2i&0
\end{array} \right)\, ,
\quad (\gamma^{\bar w})^{+}_{-} = -2i\,,\quad
\gamma^{\bar w}_{--} = -i\,.
\en

%%%%%%%%%%%%%%%%%%%%%%%%%%%%%%%%%%%%%%%%%%%%%%%%%%%%%%%%%%%%   SUPERSYMMETRY           %%%%%%%%%%%%%%%%%%%%%%%%%
\subsection{Supersymmetry transformations}
We now assume that the system at hand is endowed with an action of local  supersymmetry
transformations
$\delta_{\hat\epsilon}$ labeled by fermionic real spinor fields 
$\hat\epsilon^{r}(x,\tau)$,
antiperiodic in $\tau$.
These are the superpartners of the ordinary deformations
of space-time.
The transformations satisfy the algebra
\eq
\label{susy_alg}
[\delta_{\hat\epsilon_{1}}, \delta_{\hat\epsilon_{2}}] = 2
\hat\epsilon_{1}^{r}\hat\epsilon_{2}^{s}\gamma^{\mu}_{rs}\partial_{\mu}
\, .
\en
The vector fields on the right hand side of
(\ref{susy_alg}) must preserve the boundary,
which requires a condition $\hat\epsilon^{+}=\pm \hat\epsilon^{-}$ on the boundary.
The choice of sign is conventional.  We adopt
\begin{equation}\label{bc}
\hat\epsilon^{+}(0,\tau) = \hat\epsilon^{-}(0, \tau)\equiv \hat\epsilon(\tau) \, .
\end{equation}

The supersymmetry transformations are generated by a local fermionic 
current $G_{\mu r}\tot $ whose Ward identities are
\begin{equation} \label{Wardgeneral}
\langle \delta_{\hat\epsilon}\mathcal{O}\rangle = \iint dxd\tau \; \partial^{\mu}\hat\epsilon^{r}
\expvalc{G_{\mu r}\tot (x,\tau) \mathcal{O} }
\end{equation}
 where $\mathcal{O}$ stands for an arbitrary 
insertion of local operators and the spinor field 
$\hat\epsilon^{r}(x,\tau)$ vanishes at large $x$.
The operator  $G_{\mu r}\tot (x,\tau)$ in the above expression is understood  as 
a distribution on the half-cylinder that can have singularities on the boundary and at the 
points of insertion of other local operators.
Choosing $\hat\epsilon^{r}$ to vanish near the insertions we obtain the conservation equation 
\begin{equation}
\label{eq:superconserve}
\partial^{\mu}G_{\mu r}\tot (x,\tau)=0
\end{equation}
where the derivative is taken in the distributional sense. 

The Ward identity (\ref{Wardgeneral}) implies that the 
system with boundary is invariant under a
single global supersymmetry transformation $\mathcal{O}\rightarrow 
\mathcal{O}+\hat\epsilon\delta\mathcal{O}$ that is
generated by a conserved fermionic supercharge
\eq
\hat\epsilon\delta \mathcal{O}
= [i\hat\epsilon\hat Q, \,\mathcal{O}]
\en
where
\eq
\hat Q
= \int dx \;\hat\rho\tot (x,t)
\en
\eq
\partial_{t} \hat\rho\tot (x,t)  
+ \partial_{x} \hat \jmath\tot (x,t) = 0
\en
\eqa
\hat\rho\tot (x,t) &=& G_{t +}\tot (x,t)+G_{t +}\tot (x,t)\\
\hat \jmath\tot (x,t) &=& -G_{x +}\tot (x,t)-G_{x +}\tot (x,t)
\label{eq:supercurrent}
\,.
\ena
The supercharge density $\hat\rho\tot (x,t)$, the supercurrent
$\hat \jmath\tot (x,t)$, and the 
supercharge $\hat Q$ are all self-adjoint operators.
To derive explicitly the conservation of $\Qhat$
and the global supersymmetry transformation it generates,
substitute in the Ward 
identity a general spinor field $\hat\epsilon^{r}(x,\tau)$
that is constant in $x$
and obeys the boundary condition (\ref{bc}).
This yields, in particular, the result
\eq
\expval{i\hat Q(\tau)\, \hat\phi_{a}(0)}
= \frac12 \mathrm{sign}(\tau) \, \delta\hat\phi_{a}(0)
= \frac12 \mathrm{sign}(\tau) \, \anticomm{i\hat Q}{\hat\phi_{a}(0)}
\en
for $\hat\phi_{a}(\tau)$ a fermionic  operator localized on the 
boundary.  The right hand side is the unique solution of the Ward 
identity anti-periodic in $-\beta/2 \le \tau \le \beta/2$.

The bosonic stress-energy tensor satisfies
the Ward identity
\eq
\label{Wardbosonic}
\expval{v^{\mu}\partial_{\mu}\mathcal{O}} =
\iint dx d\tau \; \partial^{\mu}v^{\nu} \expvalc{
T_{\mu\nu}\tot (x,\tau) \, \mathcal{O}}
\en
from which we get
\eq
\partial_{t}\mathcal{O} =
[iH,\, \mathcal{O}]
\en
with hamiltonian
\eq
H = \int dx \; T_{tt}\tot (x,t)
\,.
\en
Consistency of
the supersymmetry algebra (\ref{susy_alg})
and the two Ward identities requires
$G_{\mu r}\tot $ and $T_{\mu\nu}\tot $
to be superpartners:
\eq
\label{tensors}
\delta_{\hat\epsilon}G_{\mu r}\tot (x,\tau)
= -2 \hat\epsilon^{s} \gamma^{\nu}_{rs} T_{\mu\nu}\tot (x,\tau)
\,.
\en
The global variations are
\eq
\anticomm{\hat Q}{G_{\mu +}\tot} = -2 T_{\mu w}\tot  \qquad
\anticomm{\hat Q}{ G_{\mu -}\tot } = 2 T_{\mu \bar w}\tot 
\,.
\label{eq:globalstressvariation}
\en
In particular, the global variation of the supercharge density gives the 
energy density,
\eq
\anticomm{\hat Q}{ \hat\rho\tot (x,t) } = 2 T_{tt}\tot (x,t)
\,
\en
implying the supersymmetry operator algebra
\eq
\hat Q^{2} = H
\en
which is consistent with the global transformation algebra
$\delta^{2} \mathcal{O} = i\partial_{t} 
\mathcal{O}$ that follows from (\ref{susy_alg}).

%%%%%%%%%%%%%%%%%%%%%%%%%%%%%%%%%%%%%%%%%%%%%%%%%%%%%%%%%%%%%%%%%%%%%%%%%%%%%%
\subsection{Bulk superconformal invariance}
A theory that is superconformal in the bulk
satisfies the operator equation
\begin{equation} \label{superconf}
(\gamma^{\mu})^{r}_{s}G_{\mu r}\tot (x,\tau) = 0\, ,  \qquad x>0
\,.
\end{equation}
We write, in the bulk,
\eq
G_{\mu r}\tot (x,\tau) =  G_{\mu r}\bulk (x,\tau)\, , \qquad x>0
\,.
\en
The bulk superconformal equation reads,
in complex coordinates,
\begin{equation}\label{superconf2}
G_{\bar w +}\bulk (x,\tau) = G_{w -}\bulk (x,\tau)=0  \, .
\end{equation}
By (\ref{tensors}), the bulk superconformal condition implies the ordinary
conformal invariance condition for the 
bulk stress-energy tensor, $T_{\mu}^{\mu}(x,\tau) = 0$, $x>0$.
The conservation law for
the nonvanishing bulk currents is
\eq
\partial_{\bar w}G_{w+}\bulk =\partial_{w}G_{\bar w -}\bulk =0
\en
so they are holomorphic and 
antiholomorphic respectively.
They are related to the conventional 
superconformal currents by
\eq
G_{w+}\bulk (w) = \frac{e^{\pi i/4}}{2\pi} G(-iw) \, , \qquad
G_{\bar w -}\bulk (\bar w) = \frac{e^{-\pi i/4}}{2\pi} \bar G(i\bar w) 
\,.
\en
The conventional superconformal currents are adapted to the alternate
quantization, called the \emph{bulk quantization},
in which $-x$ is the euclidean time coordinate,
$\tau$ is the spatial coordinate,
and $-iw = \tau-ix$ is the complex coordinate.
This rotation by $\pi/2$ is responsible for
the factors of $(-i)^{\pm 3/2}$ in the relation between the 
spin-3/2 superconformal currents.

Bulk superconformal invariance implies in addition
that the currents decay at spatial infinity as
\begin{equation} \label{asympt}
G_{\mu r}\bulk (x,\tau) \sim \exp(-3\pi x/\beta)
\qquad  x\rightarrow \infty
\,.
\end{equation}
This is equivalent to the superconformal condition
$G_{-1/2}|0\rangle = \bar G_{-1/2}|0\rangle =0$
on the bulk ground state $|0\rangle$ at $x=\infty$ in the bulk 
quantization.
The operators $G_{-1/2}$, $\bar G_{-1/2}$ 
are the usual Fourier modes of $G(-iw)$ and $\bar G(i\bar w)$ respectively. 
The bulk ground state is the only state in the bulk 
quantization that contributes at large $x$
in the limit where the bulk system is infinitely long,
$L/\beta \rightarrow \infty$.

%%%%%%%%%%%%%%%%%%%%%%%%%%%%%%%%%%%%%%%%%%%%%%%%%%%%%%%%%%%%%%%%%%%%%%%%%%%%%%%%%%%%%%%%%
\subsection{The boundary supercharge}
When the bulk system is superconformally invariant, the chirality of the bulk currents
$G_{w+}\bulk $, $G_{\bar w -}\bulk $ ensures that they stay
finite on the boundary.\footnote{For a non-conformal bulk theory,
a blow-up in the bulk supercurrent $G_{\mu 
r}\bulk $ at the boundary
would be compensated
by subtractions in the construction of the total distributional current $G_{\mu r}\tot $.}
The total current can be written
\begin{equation}\label{Gexp}
G_{\mu r}\tot (x,\tau) = G_{\mu r}\bulk (x,\tau) - \hat \theta_{\mu r}(\tau)\delta(x) \, .
\end{equation} 
Boundary terms proportional to 
derivatives of $\delta(x)$ are excluded
by our assumption that the system has no boundary operators of negative ultraviolet 
scaling dimension.

Substituting the expansion (\ref{Gexp}) into the Ward identity 
(\ref{Wardgeneral}) and integrating by parts,
we derive the boundary conservation equations
\eqa
\hat \theta_{x r}(\tau)&=&0\\
\label{cons}
\partial_{\tau}[\hat\theta_{\tau +} (\tau) +\hat\theta_{\tau -} (\tau) ]
&=&
G_{x+}\bulk (0,\tau)  +  G_{x-}\bulk (0,\tau)
\,.
\ena
%
%The Ward identity (\ref{Wardgeneral}) takes the form 
%\begin{eqnarray} \label{Wardgeneral2}
%-\langle \delta_{\hat\epsilon}\mathcal{O}\rangle &&= \iint dxd\tau \; \partial^{\mu}\hat\epsilon^{r}\langle
%G_{\mu r}(x,\tau) \mathcal{O} \rangle -2
%\int  d\tau \; \partial_{\tau}\hat\epsilon(\tau) \langle \hat \theta(\tau) \mathcal{O}  \rangle %\nonumber \\
%&&-\int d\tau\; \partial_{x}\hat\epsilon^{r}\langle [ \hat \theta_{x r}(\tau) -\partial_{\tau}\hat \theta^{(1)}_{\tau r}(\tau)]  \mathcal{O}  \rangle \, . 
%\end{eqnarray}
%
It is convenient to introduce the operators
\eqa
\hat \theta &=& \frac{i}{2}(\hat \theta_{\tau +}+\hat \theta_{\tau -})
= \frac12(\hat \theta_{t +}+\hat \theta_{t -}) \\
\hat q &=& -2 \hat\theta
\, .
\ena
The boundary conservation equation now reads
\eq
-2i \partial_{\tau}\hat\theta (\tau)
= G_{x+}\bulk (0,\tau)  +  G_{x-}\bulk (0,\tau)
\en
or, switching to real time,
\eq
\partial_{t}\hat q(t) + \hat \jmath\bulk (0,t) = 0
\en
%NEW-2008.09.22
where $\hat q(t)=-2\hat\theta(t)$ is the boundary supercharge.
The supercharge density and supercurrent
are separated into bulk and boundary parts
\eqa
\hat\rho\tot (x,t) &=& \hat q(t)\delta(x) + \hat\rho\bulk (x,t)  \nonumber \\
\hat \jmath(x,t) &=& \hat \jmath\bulk (x,t)
\ena
and the bulk parts are written in terms of the chiral currents
\eqa
\hat\rho\bulk (x,t) &=& G_{t +}\bulk (x,t) + G_{t -}\bulk (x,t)  \nonumber \\
\hat \jmath\bulk (x,t) &=& -G_{x +}\bulk (x,t) - G_{x -}\bulk (x,t)\,.
\ena

The stress-energy tensor
%NEW-2008.09.22 END
is obtained by varying the supercurrent, equation (\ref{tensors}),
so it takes the form
\eq\label{eq:theta}
T_{\mu\nu}\tot (x,\tau)= T_{\mu\nu}\bulk (x,\tau) - \theta_{\mu\nu}(\tau)\delta(x)
\en
where the only nonvanishing boundary component is 
$\theta_{\tau\tau}$.
Again, it is convenient to introduce
\eq
\theta(\tau) = - \theta_{\tau\tau}(\tau) = \theta_{tt}(\tau)
\en
so the boundary energy is $-\theta(t)$.

Because of the bulk conformal invariance,
the trace of the stress-energy tensor lives entirely in the boundary
\eq
T_{\mu}^{\mu}(x,\tau) = \theta(\tau) \delta(x)
\en
so $\theta(\tau)$ expresses the departure from conformal 
invariance in the system with boundary.\footnote{This formula motivates
the choice of sign in equation (\ref{eq:theta}) defining 
$\theta(\tau)$.}
>From (\ref{eq:globalstressvariation}) we see that
the operators $\hat\theta(\tau)$ and $\theta(\tau)$ are superpartners:
\eq
\delta \hat\theta(\tau) = i  \theta(\tau)\, , 
\qquad \anticomm{\hat Q}{ \hat\theta(t)}=\theta(t)
\,.
\en

We choose a complete set $\{\hat\phi_{a}(\tau)\}$ of self-adjoint
fermionic boundary operators.
Their self-adjoint superpartners are the bosonic boundary operators
$\phi_{a}(\tau)$,
\eq
\delta \hat\phi_{a}(\tau) = i \phi_{a}(\tau)\, , 
\qquad \anticomm{\hat Q}{ \hat\phi_{a}(\tau)}=\phi_{a}(\tau)
\, .
\en
The space of supersymmetric boundary conditions is parameterized  by
the boundary coupling constants $\lambda^{a}$ coupled to the
$\phi_{a}(\tau)$ as in equation (\ref{ccs}).
These couplings preserve supersymmetry because
\begin{equation}
\delta \phi_{a}(\tau) =  i  \partial_{\tau}\hat \phi_{a}(\tau)
\end{equation}
so the variation of the lagrangian is a total derivative in time.

Expanding $\hat\theta(\tau)$ in the complete set of fermionic boundary 
operators,
\eq
\hat\theta(\tau) = \beta^{a}\hat\phi_{a}(\tau)
\en
so
\eq
\theta(\tau) = \beta^{a}\phi_{a}(\tau)
\en
so the coefficients $\beta^{a}$ are the boundary beta-functions.
The entire system becomes superconformally invariant when 
$\hat\theta(\tau)$ vanishes,
given the bulk superconformal invariance.
Then, from (\ref{cons}), the boundary conservation equation
becomes $e^{3\pi i/4}G = e^{-3\pi i/4}\bar G$,
in terms of the conventional superconformal currents,
which is the standard superconformal gluing condition on the cylinder.

In proving the gradient formula, we will use correlation 
functions and response functions of the boundary supercharge $\hat q(\tau)$
and the bulk currents $G_{\mu r}\bulk(x,\tau)$.
We suppose that the correlation functions of the physical currents
$G_{\mu r}(x,\tau)$ are given.
We can define the correlation functions of $\hat q(\tau)$ by an approximation such as
\eq
\hat q_{\epsilon}(\tau)
=  \int_{0}^{\epsilon}dx \; \hat \rho(x,\tau)\, ,
\en
\eq
\expval{\hat q(\tau)\,\mathcal{O}}
= \lim_{\epsilon\rightarrow 0} 
\expval{\hat q_{\epsilon}(\tau)\,\mathcal{O}}
\,.
\en
The approximation can be controlled by virtue of the bulk 
superconformal invariance and the consequent chirality of the bulk 
currents,
\eq
\expval{\hat q_{\epsilon'}(\tau)\,\mathcal{O}}
-\expval{\hat q_{\epsilon}(\tau)\,\mathcal{O}}
= \int_{\epsilon}^{\epsilon'}dx \; \expval{ 
[G_{w+}\bulk(x,\tau)+G_{\bar w-}\bulk(x,\tau)]\,\mathcal{O}}
\,.
\en
Canonical UV behavior at the boundary 
ensures that the correlation functions of
$\hat q_{\epsilon}(\tau)$
exist in the limit and are independent of the method of 
approximation, up to a limited set of possible contact terms in $\tau$.
The boundary supercharge $\hat q(\tau)$, so defined, then differs 
within 
correlation functions
from the linear combination $\beta^{a}\hat\phi_{a}(\tau)$ of physical 
boundary operators by a similarly limited set of possible contact 
terms in $\tau$.
We need only ensure that our calculations are insensitive to these
limited sets of possible contact terms.

%%%%%%%%%%%%%%%%%%%%%%%%%%%%%%%%%%%%%%%%%%%%%%%%%%%%%%%%%%  THE PROOF %%%%%%%%%%%%%%%%%%%%%%%%%%%%%

\section{Proof of the gradient formula using Euclidean field theory}
\label{sect:euclidean_proof}
\setcounter{equation}{0}

We assume that our supersymmetric 1-D system with boundary
is unitary and is superconformally invariant in the bulk.
We also assume some regularity in the short distance behavior
at and near the boundary.
We require the following limits 
to exist (in the distributional sense),
\eq
\lim_{\epsilon\rightarrow 0} \epsilon \langle \hat \phi_{a}(\epsilon 
\tau)\hat\phi_{b}(0)\rangle, \quad
\lim_{\epsilon\rightarrow 0} \epsilon \langle \hat \phi_{a}(\epsilon 
\tau)\hat\theta(0)\rangle , \quad
\lim_{\epsilon\rightarrow 0} \epsilon^{2}
\langle \hat G_{\mu r}\bulk (\epsilon x, \epsilon \tau)\hat\phi_{a}(0)\rangle 
\,,
\en
and we require that there be no operators whose UV scaling dimension is negative.
These requirements on the short distance behavior
are satisfied if there is a supersymmetric short distance fixed point of the RG 
(thereby permitting canonical scaling analysis),
and if the UV fixed point theory
satisfies a weak cluster decomposition principle,
that correlation functions should not grow 
at large separation (thereby forbidding negative dimension operators).
Our short distance assumptions imply constraints on the contact terms 
that can occur in boundary correlation functions:
\eq\label{canonicalUV2}
\lim\limits_{\epsilon\to 0} \int\limits_{|\tau|<\epsilon}d\tau \, \tau^{k}
\langle \hat \phi_{a}(\tau)\hat\phi_{b}(0)\rangle = 
\lim\limits_{\epsilon\to 0} \int\limits_{|\tau|<\epsilon}d\tau \, \tau^{k}
\langle \hat \phi_{a}(\tau)\hat\theta(0)\rangle =0\,,
\quad \mbox{for $k\ge 1$}\, .
%\lim\limits_{\epsilon\to 0} \iint\limits_{|\tau^{2} +x^{2}|<\epsilon}
%(\tau^{2}
% + x^{2})^{k} \langle \hat G_{\mu r}(x, \tau)\hat\phi_{a}(0)\rangle &=& 0 
%\, , \enspace \mbox{for any $k\ge 1$}\, . \mbox{ do I need the last one?}
\en

The Ward identities for conformal Killing spinor fields are of
particular interest, given bulk superconformal invariance.
A spinor field $\hat\epsilon^{r}(x,\tau)$ is a conformal Killing spinor field if there
exists a spinor field $\eta_{s}(x,\tau)$ such that
\begin{equation} \label{Killing}
\partial^{\mu}\hat\epsilon^{r} = (\gamma^{\mu})^{r}_{s} \hat\eta^{s}
\end{equation}
(which means that the local supersymmetry transformation generated by $\hat\epsilon^{r}$
is compensated by the superWeyl transformation generated by 
$\hat\eta^{s}$).
In complex coordinates equation (\ref{Killing}) reads
\begin{equation}
\partial^{w}\hat\epsilon^{+}= 2\partial_{\bar w}\hat\epsilon^{+} = 0 \, , \quad
\partial^{\bar w}\hat\epsilon^{-} =  2 \partial_{ w}\hat\epsilon^{-} = 0\,,\quad
\partial^{\bar w}\hat\epsilon^{+} = -4i\hat\eta_{+}
\,,\quad
\partial^{w}\hat\epsilon^{-} = 4i \hat\eta_{-}
\label{eq:Killingcomplex}
\end{equation}
So the conformal Killing condition is the condition that
the components $\hat\epsilon^{+}$ and $\hat\epsilon^{-}$ be
holomorphic and antiholomorphic respectively
(and complex conjugate to each other, to satisfy
the euclidean reality condition).

We choose a certain special conformal Killing spinor field
for each point $\tau'$ on the boundary:
\eq
\hat\epsilon^{+}(w) =  
\hat\epsilon_{0}
\cosh\left [\frac{\pi(w-i\tau')}{\beta}\right ]
\,,
\quad
\hat\epsilon^{-}(\bar w) =  
\hat\epsilon_{0}
\cosh\left [\frac{\pi(\bar w + i\tau')}{\beta}\right ]
\label{ksp}
\en
where $\hat\epsilon_{0}$ is an arbitrary real fermionic constant.
This special spinor field $\hat\epsilon^{r}(x,\tau)$ is antiperiodic in $\tau$, 
satisfies the conformal Killing constraints (\ref{eq:Killingcomplex}) 
with
\eq
\hat\eta_{+} = \hat\epsilon_{0}\eta(w-i\tau')\,,\quad
\hat\eta_{-} = \hat\epsilon_{0}\bar\eta(\bar w+i\tau')\,\quad
\eta(w) = \frac{i\pi}{2\beta} \sinh
\left (\frac{\pi w}{\beta}\right )\,,
\label{eq:eta}
\en
and satisfies the boundary condition (\ref{bc})
with boundary spinor field
\begin{equation}\label{bspin}
\hat\epsilon(\tau) =   \hat\epsilon_{0} \cos \left [ 
\frac{\pi(\tau-\tau')}{\beta} \right] \, .
\end{equation}

Let us consider the Ward identity (\ref{Wardgeneral}) corresponding to this 
special conformal spinor field, with the insertion of a single 
boundary fermion field $\hat 
\phi_{a}(\tau')$,
\begin{equation} \label{spWard}
\langle \delta_{\hat\epsilon}\hat\phi_{a}(\tau') \rangle
=\iint  dxd\tau \; \partial^{\mu}\hat\epsilon^{r}(x,\tau) \langle
G_{\mu r}\tot (x,\tau) \hat \phi_{a}(\tau') \rangle 
\,.
\end{equation} 
Even though the special spinor field $\hat\epsilon^{r}$ blows up at large $x$,
it can be used in the Ward identity
because of the asymptotic condition (\ref{asympt}) that follows from 
superconformal invariance of the bulk ground state.
We can substitute on the left hand side the global variation 
\eq
\langle \delta_{\hat\epsilon}\hat\phi_{a}(\tau') \rangle 
= \hat\epsilon(\tau')\langle \delta\hat\phi_{a}(\tau')\rangle
= 
i\hat\epsilon_{0}\langle\phi_{a}\rangle
\en
because the first derivatives $\partial_{\mu}\hat\epsilon^{r}$ of our special spinor field vanish at the insertion 
point, and because any higher derivative contributing to 
$\delta_{\hat\epsilon}\hat\phi_{a}$ would have a negative 
dimension boundary operator as coefficient.
By translation invariance in $\tau$ we can choose $\tau'=0$ in the 
Ward identity (\ref{spWard}) without loss of generality.
Substituting (\ref{Gexp}) into (\ref{spWard}), using 
the conformal Killing property (\ref{Killing}) and dropping the common 
factor $i\hat\epsilon_{0}$ we obtain  
\begin{eqnarray}
\langle \phi_{a} \rangle
&=& \iint  dxd\tau \; 
\left [
4 \eta(\bar w) \langle  G_{w -}\bulk (x,\tau)\, \hat \phi_{a}(0) \rangle  
- 4 \eta(w)\langle  G_{\bar w +}\bulk (x,\tau)\, \hat \phi_{a}(0) \rangle  
\right ]
\nonumber \\
&& +  
\int   d\tau \;
4 \eta(i\tau)
\left [
\langle \hat \theta(\tau)\, \hat \phi_{a}(0) \rangle  
+
\langle \frac{1}{2}(\hat \theta_{x-}(\tau)-\hat \theta_{x+}(\tau)) \,
\hat \phi_{a}(0) \rangle  
\right ]
\, .
\end{eqnarray}
Taking into account the explicit form (\ref{eq:eta}) of $\eta(w)$ 
we get
\begin{equation}\label{gradientwithE}
\langle \phi_{a} \rangle =
\frac{1}{\beta} \frac{\partial \ln z}{\partial \lambda^{a}}
=
E - \frac{2\pi}{\beta}\int_{0}^{\beta}   d\tau \, 
\sin ({\pi\tau}/{\beta} ) \langle \hat \theta(\tau)
\hat \phi_{a}(0) \rangle
\, .
\end{equation}
where
\begin{eqnarray}\label{E}
E  = &&
\iint  dxd\tau \; 
\left [
4 \bar \eta(\bar w)\langle G_{w -}\bulk (x,\tau)\hat \phi_{a}(0) \rangle
 - 4\eta(w)\langle G_{\bar w +}\bulk (x,\tau)\hat \phi_{a}(0) \rangle 
\right ]
\nonumber \\
&& +  
\int   d\tau \; 2 \eta(i\tau)\left [ \langle 
\hat \theta_{x-}(\tau)\hat \phi_{a}(0) \rangle 
-\langle \hat \theta_{x+}(\tau)\hat \phi_{a}(0) \rangle    
\right ]
 \, .
\end{eqnarray}

We now argue that the quantity $E$ vanishes under the assumptions on UV behavior.
The correlation functions of the
bulk currents $G_{\bar w +}\bulk (x,\tau)$, $G_{w -}\bulk (x,\tau)$ 
vanish up to contact terms,
because of the bulk conformal invariance (\ref{superconf2}).
Thus the  two point functions in the first line of (\ref{E})  
are linear combinations of  $\delta(x)\delta(\tau)$ and its
derivatives\footnote{There are no terms of
the form $f(\tau)\delta(x)$ where $f$ is a smooth function because
the supercurrent has been split into bulk and boundary parts
so that such terms are all contained
in the $\langle\hat \theta(\tau)\hat \phi_{i}(0)\rangle$ correlators.} 
%If the UV behavior of the theory is
%governed by some fixed point the boundary operators $\phi_{a}$ have the UV scaling %dimension at most
%one and therefore that of the fields $\hat \phi_{a}$ is at most $1/2$. The bulk %supercurrent
%$G_{\mu r}$ has a canonical scaling dimension $3/2$. 
The assumptions on UV behavior then imply  that the correlators
$\langle G_{\bar w + }\bulk (x,\tau) \hat \phi_{a}(0) \rangle$,
$\langle G_{ w - }\bulk (x,\tau) \hat \phi_{a}(0) \rangle$ are each proportional
to $\delta(x)\delta(\tau)$.
There are no higher order contact terms.
Such terms however vanish upon integration in (\ref{E}) because the 
functions $\eta(w)$,
$\bar \eta(\bar w)$ vanish at the insertion point $x=0$, $\tau=0$.
Therefore the term in the first line in (\ref{E}) 
vanishes. The terms in the second line contain the operators $\hat \theta_{x\pm}$
that vanish by the equations of motion (\ref{cons}),
so their correlators are pure contact terms.
It follows from  (\ref{canonicalUV2}) that the
contact terms in the correlators in the second line of $E$ can
be no more singular than $\delta(\tau)$,
and hence  vanish upon integration with $\eta(i\tau)$,
which vanishes at $\tau=0$.
Therefore $E=0$.

Next, we substitute
$\beta^{a}\hat \phi_{a}$ for $\hat \theta$ in (\ref{gradientwithE}).
The canonical UV behavior (\ref{canonicalUV2}) makes this possible.
The correlation function might be changed by a contact
term, but nothing more singular than $\delta(\tau)$.   The smearing function
$\sin (\pi\tau/\beta)$ vanishes at $\tau=0$ so such a
contact term would have no effect.\footnote{A similar step is implicitly
present in the proof of bosonic gradient formula given in \cite{FK}.}
We obtain the gradient formula
\begin{equation}
\frac{\partial \ln z }{\partial \lambda^{a}} = -\gsusy_{ab}\beta^{b}
\end{equation}
with
\begin{equation}
\label{eq:metriceuclidean}
\gsusy_{ab}=
2\pi \int_{0}^{\beta}d\tau \, \sin(\pi \tau/\beta ) \langle
\hat \phi_{a}(\tau) \hat \phi_{b} (0) \rangle
\,.
\end{equation}
%where the metric $g_{ab}^{\rm N=1}$ is given by formula (\ref{susymetric}).
%\begin{equation}
%g_{ab}^{\rm N=1} = \pi\! \int\!   d\tau \, \sin[\pi(\tau)/\beta] \langle
%\hat \phi_{a}(\tau) \hat \phi_{b} (0) \rangle \, .
%\end{equation}
% At the last step in (\ref{gf}) we used the relation (\ref{hat_theta}) that holds up to contact terms.
% By the canonical UV behavior  (\ref{canonicalUV2}) 
% such contact terms between $\hat \phi_{a}$ and $\hat \phi_{b}$ and
% between $\hat \phi_{a}$ and $\hat \theta$ can be only proportional to $\delta(\tau-\tau')$ and therefore
% can be neglected as the smearing function $\sin[\pi(\tau)/\beta]$ vanishes at coincident 
% points\footnote{A similar step is implicitely present  in the proof of bosonic gradient formula given in \cite{FK}.}. 
To see that the metric $\gsusy_{ab}$ is positive-definite, we rewrite it
\begin{equation} \label{metricpositive}
\int_{0}^{\beta} d\tau \; \sin (\pi\tau/\beta) \langle
\hat \phi_{a}(\tau) \hat \phi_{b} (0) \rangle = 
\lim\limits_{\epsilon\to 0} 
\int_{\epsilon}^{\beta-\epsilon} d\tau \; \sin(\pi\tau/\beta)\langle
\hat \phi_{a}(\tau) \hat \phi_{b} (0) \rangle \,,
\end{equation}
again making use of the canonical UV behavior (\ref{canonicalUV2}).
The operators $\hat \phi_{a}$ are self-adjoint, so the two-point function
at finite separation is positive by reflection positivity.
Therefore the right hand side of (\ref{metricpositive}) is positive.

% The only threat to positivity are contact terms that may be present in 
%the two point functions. The only contact term allowed in 
%$\hat \phi_{a}(\tau) \hat \phi_{b} (\tau') \rangle $ by (\ref{canonicalUV2}) is 
%$\delta(\tau-\tau')$. It drops out after integration with the smearing function. 

%The above proof  relies on the following set of assumptions: local conservation of %supersymmetry 
%(\ref{cons}), superconformal invariance of the bulk theory (\ref{superconf}), %(\ref{asympt}), 
%renormalizability (\ref{hat_theta}), existense of UV fixed point and global invariance %under Euclidean time translations.
The proof depends on the canonical UV behavior at three points: 
the vanishing of the term $E$ in (\ref{gradientwithE}),
the substitution of $\beta^{a}\hat \phi_{a}$ for $\hat \theta$,
and the positivity of the metric.
The issue 
in all three cases is that operator identities apply in 
correlation functions only up to contact terms.
The technique of the present proof is a subtle improvement on the 
proof for the general gradient formula \cite{FK}.
There we used the bulk and boundary conservation equations separately.
Here we use the single Ward identity (\ref{Wardgeneral}).
This is  more economic and also more transparent as we do not need to worry about
the contact terms associated with the separate conservation equations.
In essence the above euclidean proof hinges on the special Ward
identity plus the assumptions about canonical UV behavior.

% Note that  basic equations (\ref{cons}), (\ref{superconf}), (\ref{hat_theta})  are all operator 
% equations holding up to contact terms. While independence of the gradient formula on the contact terms
%  related to (\ref{superconf}), (\ref{hat_theta}) is ensured by the short distance 
% canonical behavior  assumption (\ref{canonicalUV2}), 
% the contact terms present in the two equations in (\ref{cons}) are constrained by the Ward identity 
% (\ref{Ward}). It is essential for the sake of the proof that we use Ward identity (\ref{Ward})
%  rather than  using the bulk and boundary conservation   equations  (\ref{cons}) in two separate steps.
% The derivation of the general gradient formula \cite{FK} can be recast in a similar way. Instead of using 
% the bulk and boundary stress-energy tensor conservation equations one can manipulate the 
% whole Ward identity used in \cite{FK}. 
% %[For rigid supersymmetry transformations, that is for constant $\hat\epsilon^{a}$,
% %this variation does not vanish due to the antiperiodicity of $G_{\mu a}$ and $\hat \theta$ in the
% %$\tau$ direction. This reflects the spontaneous symmetry breaking of rigid supersymmetry in the NS %sector.]

%%%%%%%%%%%%%%%%%%%%%%%%%%%%%%%%%%%%%%%%%%%%%%%%%%%%%%%%%%%%%%%%%%%%%%%%%%%%%%%%%%%%%%%%%%%%%%%%%%%%%%%%%%%%%
%%%%%%%%%%%%%%%%%%%%%%%%%%%%%%        REAL    TIME     METHODS                           %%%%%%%%%%%%%%%%%%%%
%%%%%%%%%%%%%%%%%%%%%%%%%%%%%%%%%%%%%%%%%%%%%%%%%%%%%%%%%%%%%%%%%%%%%%%%%%%%%%%%%%%%%%%%%%%%%%%%%%%%%%%%%%%%%
\section{Proof of the gradient formula using real time field theory}
\label{sect:realtime_proof}
\setcounter{equation}{0}

Here we give a second proof of the gradient formula (\ref{susygradf}),
using real time methods to evaluate
\eq
\frac{\partial\ln z}{\partial \lambda^{a}} 
= \beta \expval{\phiaB} \\
= \beta \expval{\anticomm{\Qhat}{\phiaF}}
\,.
\en
First, we
separate the supercharge into 
the contribution $\hat q_{\epsilon}(t)$ from a neighborhood of the boundary
and the contribution $\Qhat_{\epsilon}(t)$ from the rest of the system:
\eq
\hat q_{\epsilon}(t) = \int_{0}^{\epsilon} dx \; \hat \rho(x,t) 
\qquad
\Qhat_{\epsilon}(t) = \hat Q - \hat q_{\epsilon}(t)
\,.
\en
Let
\eqa
f_{a,\epsilon}(\omega) &=& \int_{-\infty}^{\infty} dt\; e^{i\omega 
t}\expval{\anticomm{\hat q_{\epsilon}(t)}{\phiaF(0)}} \\
F_{a,\epsilon}(\omega) &=& \int_{-\infty}^{\infty} dt\; e^{i\omega 
t} \expval{\anticomm{\hat Q_{\epsilon}(t)}{\phiaF(0)}}
\label{eq:bigF}
\ena
so that
\eq
\label{eq:split}
2\pi \delta(\omega) \expval{\phiaB}
= f_{a,\epsilon}(\omega) + F_{a,\epsilon}(\omega)
\,.
\en
It is convenient to introduce an IR regulator $\delta>0$ into equation (\ref{eq:bigF}),
\eq\label{eq:bigFreg}
F_{a,\epsilon}(\omega) =
\lim_{\delta\rightarrow 0}\int_{-\infty}^{\infty} dt\; e^{i\omega 
t-\delta|t|} \expval{\anticomm{\hat Q_{\epsilon}(t)}{\phiaF(0)}}
\,,
\en
in order to regularize the singularity at $\omega=0$ in intermediate 
stages of our calculation.

Locality tells us that,
for $t$ sufficiently near $0$,
\eq
\anticomm{\hat Q_{\epsilon}(t)}{\phiaF(0)}
= 0
\,.
\en
We combine this with charge conservation at $x=\epsilon$,
\eq
\partial_{t} \hat Q_{\epsilon}(t) = \hat \jmath\bulk(\epsilon,t)
\, ,
\en
to get the identity
\eq
\label{eq:intformula}
\anticomm{\hat Q_{\epsilon}(t)}{\phiaF(0)}
= \int_{0}^{t} dt'\; \anticomm{\hat \jmath\bulk(\epsilon,t')}{\phiaF(0)}
\;.
\en
We use this identity in (\ref{eq:bigFreg})
to derive
\eqa
\label{eq:Faepsilon}
F_{a,\epsilon}(\omega) &=&
\lim_{\delta\rightarrow 0}
\left [
\frac{R^{+}_{a,\epsilon}(\omega)}{\omega + i\delta} 
+ \frac{R^{-}_{a,\epsilon}(\omega)}{\omega - i\delta}
\right ] \nonumber \\
&=& i \pi \delta(\omega)\left [R^{-}_{a,\epsilon}(0)- R^{+}_{a,\epsilon}(0)\right ]
+ \mathcal{P}(1/\omega)\left [R^{+}_{a,\epsilon}(\omega)+R^{-}_{a,\epsilon}(\omega)\right ] 
\ena
where $R^{\pm}_{a,\epsilon}(\omega)$ are the response functions
\eq
R^{\pm}_{a,\epsilon}(\omega)
= \pm \int_{0}^{\pm \infty}dt\; e^{i\omega t } 
\expval{\anticomm{i \hat \jmath\bulk(\epsilon,t)}{\phiaF(0)}}
\,.
\en
We do without the IR regulator $\delta$ in the construction of the 
response functions, because they are regular at $\omega=0$,
otherwise the correlation functions $F_{a,\epsilon}(\omega)$ would 
be more singular than $\delta(\omega)$, meaning that the real time 
correlators would grow with time.
$R^{+}_{a,\epsilon}(\omega)$ is analytic in the upper-half plane,
and 
$R^{-}_{a,\epsilon}(\omega)$ is analytic in the lower-half plane.

The bulk supercurrent separates into the two chiral superconformal currents,
\eq
\hat \jmath\bulk(x,t) = -G_{w+}\bulk(x,t) - G_{\bar w -}\bulk(x,t) \,.
\en
Chirality implies that
\eqa
G_{w+}\bulk(\epsilon,t) &=& G_{w+}\bulk(\epsilon-t,0) \qquad t < 
+\epsilon \,, \nonumber \\
G_{\bar w-}\bulk(\epsilon,t) &=& G_{\bar w-}\bulk(\epsilon+t,0) 
\qquad t > - \epsilon
\ena
so, by locality of the equal-time anti-commutators,
\eqa
\anticomm{ -i G_{w+}\bulk(\epsilon,t)}{\phiaF(0)} &=& 0 \qquad t < +\epsilon \,, \nonumber \\
\anticomm{ -i G_{\bar w-}\bulk(\epsilon,t)}{\phiaF(0)}&=& 0 \qquad t > - \epsilon
\ena
so
\eqa
\anticomm{i \hat \jmath\bulk(\epsilon,t)}{\phiaF(0)}
&=& \anticomm{ -i G_{\bar w-}\bulk(\epsilon,t)}{\phiaF(0)}
\qquad t < +\epsilon \,,\nonumber \\
\anticomm{i \hat \jmath\bulk(\epsilon,t)}{\phiaF(0)}
&=& \anticomm{ -i G_{w+}\bulk(\epsilon,t)}{\phiaF(0)}
\qquad t > - \epsilon
\,
\ena
so we can write
\eqa
R^{+}_{a,\epsilon}(\omega) &=& 
\int_{0}^{\infty}dt \; e^{i\omega t} 
\expval{\anticomm{-iG_{w +}\bulk(\epsilon,t)}{\phiaF(0)}} \nonumber \\
&=& \int_{-\infty}^{\infty}dt\; e^{i\omega t } 
\expval{\anticomm{-iG_{w +}\bulk(\epsilon,t)}{\phiaF(0)}} \, , \\
R^{-}_{a,\epsilon}(\omega) &=& 
\int_{-\infty}^{0}dt\; e^{i\omega t } 
\expval{\anticomm{-iG_{\bar w -}\bulk(\epsilon,t)}{\phiaF(0)}} \nonumber \\
&=& 
\int_{-\infty}^{\infty}dt\; e^{i\omega t } 
\expval{\anticomm{-iG_{\bar w -}\bulk(\epsilon,t)}{\phiaF(0)}}
\,.
\ena
The dependence on $\epsilon$ is trivial
because of the chirality, now in the form
\eq
G_{w+}\bulk(\epsilon,t) = G_{w+}\bulk(0,t-\epsilon)\,, \qquad
G_{\bar w-}\bulk(\epsilon,t) = G_{\bar w-}\bulk(0,t+\epsilon) \,.
\en
We have
\eq\label{eq:epsdep}
R^{+}_{a,\epsilon}(\omega) = e^{+ i\omega\epsilon } R^{+}_{a}(\omega)
\,,
\qquad
R^{-}_{a,\epsilon}(\omega) = e^{- i\omega\epsilon } R^{-}_{a}(\omega)
\,,
\en
with
\eqa
R^{+}_{a}(\omega) &=& 
\int_{-\infty}^{\infty}dt\; e^{i\omega t } 
\expval{\anticomm{-iG_{w +}\bulk(0,t)}{\phiaF(0)}}
\label{eq:Rplus} \, \nonumber \\
R^{-}_{a}(\omega) &=& 
\int_{-\infty}^{\infty}dt\; e^{i\omega t } 
\expval{\anticomm{-iG_{\bar w -}\bulk(0,t)}{\phiaF(0)}}
\label{eq:Rminus}\,.
\ena
The limit $\epsilon\rightarrow 0$
of equation (\ref{eq:Faepsilon})
is now taken easily,
\eqa 
\label{eq:Fa}
F_{a}(\omega) \equiv \lim_{\epsilon\rightarrow 0} F_{a,\epsilon}(\omega) 
&=& \lim_{\delta\rightarrow 0}
\left [
\frac{R^{+}_{a}(\omega)}{\omega + i\delta} 
+ \frac{R^{-}_{a}(\omega)}{\omega - i\delta}  
\right ]\, \nonumber 
\\
&=& i\pi\delta(\omega) \left [ R^{-}_{a}(0)-R^{+}_{a}(0)\right ] 
+ \mathcal{P}(1/\omega)\left [R^{+}_{a}(\omega)+R^{-}_{a}(\omega)\right ]  .
\ena
Then, from equation (\ref{eq:split}), we get the limit
\eq
f_{a}(\omega) \equiv \lim_{\epsilon\rightarrow 0} f_{a,\epsilon}(\omega)
= 2\pi \delta(\omega) \expval{\phiaB}
- F_{a}(\omega)
\,.
\label{eq:fa}
\en
We have thus used the chirality of the bulk superconformal currents
to construct the correlation functions of $\hat q(t)
= \lim_{\epsilon\rightarrow 0} \hat q_{\epsilon}(t)$,
\eq\label{eq:fafab}
f_{a}(\omega)
=
\lim_{\epsilon\rightarrow 0} 
\int_{-\infty}^{\infty} dt\; e^{i\omega 
t}\expval{\anticomm{\hat q_{\epsilon}(t)}{\phiaF(0)}}
= \int_{-\infty}^{\infty} dt\; e^{i\omega 
t}\expval{\anticomm{\hat q(t)}{\phiaF(0)}}
\,.
\en

At this point, we could assume that $f_{a}(\omega)$ has no 
delta-function contribution at $\omega=0$, and conclude from 
(\ref{eq:fa}) that
\eq
%NEW-2008.09.22
\expval{\phiaB} = \frac{i}2 R^{-}_{a}(0)-\frac{i}2 R^{+}_{a}(0)
%NEW-2008.09.22 END
\,.
\en
This is the assumption that the 
boundary correlators decay in time, that all boundary degrees of 
freedom return to equilibrium after any perturbation in the boundary.
This is essentially the assumption that all boundary degrees of 
freedom couple to the bulk, thereby thermalizing.
This tack was taken in \cite{FK2}.
In fact, we will not need to make
this thermalization assumption to prove the gradient formula.

Our next step is to show that the global bulk superconformal invariance
expresses itself by vanishing formulas
\eq
\label{eq:vanishing}
R^{+}_{a}(i\pi/\beta) = 0\,,
\qquad R^{-}_{a}(-i\pi/\beta)=0
\,.
\en
First, we use the usual relation between thermal correlation functions and 
expectation values of anti-commutators:
\eq
\expval{\anticomm{G_{w +}\bulk(0,t)}{\phiaF(0)}} 
= \expval{G_{w +}\bulk(0,t)\,\phiaF(0)}
+ \expval{G_{w +}\bulk(0,t-i\beta)\,\phiaF(0)}
\en
to obtain
\eq
\expval{G_{w +}\bulk(x,t)\,\phiaF(0)}
= \int_{-\infty}^{\infty} \frac{d\omega}{2\pi} 
\;\frac{e^{i\omega (x-t)}}{1+e^{-\omega\beta}}
R^{+}_{a}(\omega)
\,.
\en
This expression analytically continues to
euclidean time $\tau=it$ for $0<\tau<\beta$,
\eq
\expval{G_{w +}\bulk(x,t)\,\phiaF(0)}
= \int_{-\infty}^{\infty} \frac{d\omega}{2\pi} 
\;\frac{e^{i\omega x -\omega \tau}}{1+e^{-\omega\beta}}
R^{+}_{a}(\omega)
\,.
\en
If we take $x>0$,
we can deform the contour of integration into the upper-half plane,
where the response function $R^{+}_{a}(\omega)$ is analytic.
The euclidean correlation function is then expressed as a sum of the 
residues at the thermal poles
\eq
\expval{G_{w +}\bulk(x,\tau)\,\phiaF(0)}
=
\sum_{k=1}^{\infty}
e^{- \omega_{k} (x+i \tau)} \, i \beta^{-1}R^{+}_{a}(i\omega_{k}) 
\,,
\qquad
\omega_{k} = \frac{2\pi}\beta \left ( k-\frac12 \right )
\label{eq:thermalpoles}
\,.
\en
The same thermal correlation function is given
in the bulk quantization, where $-x$ is the euclidean time,
as the matrix element
\eq
\expval{G_{w +}\bulk(x,\tau)\,\phiaF(0)}
=
\langle B |\phiaF(0)  \, G_{w +}\bulk(x,\tau) |0\rangle
\en
where $|0\rangle$ is the superconformal bulk ground state and
$\langle B |$ is the bulk state representing the boundary condition
at $x=0$.
The global superconformal invariance condition in the bulk,
$G_{-1/2} |0\rangle = 0$,
implies that the $k=1$ term vanishes in the sum (\ref{eq:thermalpoles})
over the thermal poles.
Therefore
$
R^{+}_{a}(i\pi/\beta) = 0
$.
Similarly, using the analyticity of $R^{-}_{a}(\omega)$ in the 
lower-half plane and the global bulk superconformal condition
$\bar G_{-1/2} |0\rangle = 0$, we derive the other vanishing formula
$R^{-}_{a}(-i\pi/\beta) = 0$.
The error in these vanishing formulas is exponentially small in 
$L/\beta$, the exponent given by the scaling dimension
of the most relevant operator in the bulk superconformal field 
theory.\footnote{A purely real time proof of the gradient formula
would require a real time proof of the vanishing formulas
from bulk superconformal invariance,
without appealing to the euclidean field theory.}

We can now derive a sum rule
\eqa
&&\int \frac{d\omega}{2\pi}\;
\frac{\pi^{2}/\beta^{2}}{\omega^{2} + \pi^{2}/\beta^{2}}
\; F_{a}(\omega) =
\lim_{\delta\rightarrow 0}
\int \frac{d\omega}{2\pi}\;
\frac{\pi^{2}/\beta^{2}}{\omega^{2} + \pi^{2}/\beta^{2}}
\; 
\frac{R^{+}_{a}(\omega)}{\omega + i\delta} \nonumber \\
&& %\qquad
{}+
\lim_{\delta\rightarrow 0}
\int \frac{d\omega}{2\pi}\;
\frac{\pi^{2}/\beta^{2}}{\omega^{2} + \pi^{2}/\beta^{2}}
 \frac{R^{-}_{a}(\omega)}{\omega - i\delta}
= -\frac{i}2 R^{+}_{a}(i\pi/\beta) + \frac{i}2 
R^{-}_{a}(-i\pi/\beta) =0 \, .
\label{eq:sumrule}
\ena
The calculation starts from equation (\ref{eq:Fa})
for $F_{a}(\omega)$.
In the first step, we can
exchange the integral over $\omega$ with the removal of the IR regulator
and separate the two integrals, as long as 
$R^{\pm}_{a}(\omega)/\omega^{3}$ is integrable at infinity.
Then the contours of integration are deformed into the upper and lower 
half planes, respectively.
The growth condition on $R^{\pm}_{a}(\omega)$
justifies discarding the contours at infinity.
The last step uses the vanishing formulas (\ref{eq:vanishing}).

Canonical UV behavior at the boundary guarantees that 
$R^{\pm}_{a}(\omega)$ grows at most as $\omega$,
which more than satisfies the growth condition.
The conformal supercurrents $G_{\mu r}\bulk(x,t)$ have canonical dimension $3/2$, while the 
boundary fields $\phiaF(t)$ have canonical UV dimension $1/2$.
We assume, as an aspect of the canonical UV behavior, that 
there are no negative dimension boundary operators, so no such 
operators can occur in operator products of the bulk currents and the 
boundary fields.
Therefore the response functions $R^{\pm}_{a}(\omega)$
defined by equations (\ref{eq:Rplus}), (\ref{eq:Rminus}) have 
canonical UV dimension $1$,
and can grow no faster than $\omega$ at large $\omega$.
The leeway between the canonical growth rate $\omega$ and the growth 
rate $\omega^{2}$ where the proof breaks down
allows for the possibility of
fermionic boundary fields with UV 
scaling dimensions slightly larger than $1/2$,
as in the $\alpha'\rightarrow 0$ limit of string theory.

Combining the sum rule (\ref{eq:sumrule}) with equation (\ref{eq:fa}),
we get
\eq
\frac{\partial\ln z}{\partial \lambda^{a}} 
= \beta \expval{\phiaB} = \beta \int \frac{d\omega}{2\pi}\;
\frac{\pi^{2}/\beta^{2}}{\omega^{2} + \pi^{2}/\beta^{2}}
\; f_{a}(\omega)
\,.
\en
We substitute $\hat q(t)=-2\beta^{a}\phiaF(t)$
in (\ref{eq:fafab}) to obtain
\eq
f_{a}(\omega)
= \int_{-\infty}^{\infty} dt\; e^{i\omega t}
\expval{\anticomm{-2\beta^{b}\phibF(t)}{\phiaF(0)}}
= -2 \beta^{b} f_{ab}(\omega)
\en
with
\eq
f_{ab}(\omega)
=
\int_{-\infty}^{\infty} dt\; e^{i\omega 
t}\expval{\anticomm{\phibF(t)}{\phiaF(0)}}
\,.
\en
Now we have
\eq
\frac{\partial\ln z}{\partial \lambda^{a}} 
= \beta \expval{\phiaB} = -2\beta \int \frac{d\omega}{2\pi}\;
\frac{\pi^{2}/\beta^{2}}{\omega^{2} + \pi^{2}/\beta^{2}}
\; f_{ab}(\omega)
\beta^{b}
\en
which is the gradient formula
\eq
\frac{\partial\ln z}{\partial \lambda^{a}} 
=  - \gsusy_{ab} \beta^{b}
\en
with metric
\eq
\gsusy_{ab} =
\int d\omega\;
\frac{\pi/\beta}{\omega^{2} + \pi^{2}/\beta^{2}}
\; 
f_{ab}(\omega)
\en
We can integrate out $\omega$ to get the equivalent formula
\eq
\gsusy_{ab} = \int_{-\infty}^{\infty} dt\; 
\pi e^{-\pi|t|/\beta}
\expval{\anticomm{\phibF(t)}{\phiaF(0)}}
\,.
\en

The assumption of canonical UV behavior implies
that the correlation functions $f_{ab}(\omega)$ grow
no faster than $|\omega|^{0}$,
so the metric is well-defined.
By unitarity, the $f_{ab}(\omega)$, for each $\omega$, form a non-negative hermitian 
matrix, and $f_{ab}(-\omega)= f_{ba}(\omega)$, so the metric 
$\gsusy_{ab}$ is symmetric and non-negative.  Any null vector for the 
metric,
$\gsusy_{ab}v^{a}v^{b}=0$, would be a null vector for $f_{ab}(\omega)$ 
for all $\omega$, which would imply $v^{a}\hat\phi_{a}(t)=0$, so $v^{a}=0$, 
since the $\hat\phi_{a}$ are linearly independent.
Therefore $\gsusy_{ab}$ is a positive definite metric on the space of 
boundary conditions.

Note that we have made no assumptions on the IR behavior of 
$f_{a}(\omega)$.  Equation (\ref{eq:fa}) allows for the possibility 
that $f_{a}(\omega)$ contains a long-time contribution proportional 
to $\delta(\omega)$, which is to say that the boundary energy could
fail to thermalize after a local perturbation, as when the boundary contains 
a decoupled sub-system.

The assumption of canonical UV behavior at the boundary enters the 
real time proof at several points.
We defined the correlation functions of the
boundary supercharge $\hat q(t)$
through the regularization procedure $\hat q(t)=\lim_{\epsilon\rightarrow 0}
\hat q_{\epsilon}(t)$.
We could have used some other regularized separation of the boundary from 
the rest of the system.
This could have modified $\hat q(t)$ by some boundary operator,
but that operator would have negative UV scaling dimension, which
is excluded by the assumption of canonical UV behavior.
We assumed canonical UV scaling of the correlation functions of the 
boundary fields with the bulk superconformal currents
when we derived the
superconformal sum rule (\ref{eq:sumrule}).
This requires an upper bound on the UV scaling dimensions of the 
boundary fields, and also the absence of 
negative dimension operators,
which could have nonzero expectation values at finite temperature.
Finally, we replaced $\hat\theta(t)$ by $\beta^{a}\hat \phi_{a}(t)$
in correlation functions.

The key step in the proof is the separation of the boundary from the
rest of the system by means of the sum rule
(\ref{eq:sumrule}).
Both bulk superconformal invariance
and canonical UV behavior at the boundary
are needed to derive the sum rule.
%NEW-2008.09.22
The UV regularity makes it possible to write a sum rule
if just one subtraction can be taken.
The bulk superconformal invariance
expressed in the vanishing formulas (\ref{eq:vanishing})
allows us to make that subtraction
(at a low thermal energy).
%NEW-2008.09.22 END
The bulk superconformal invariance also enters at short distance 
when the chirality of the superconformal currents is used to 
construct the boundary supercharge.
It would be good to have a physical understanding of the need for
this combination of ultraviolet and infrared technical conditions.

\section{Comparison of the two proofs}
\setcounter{equation}{0}
We should check that
the two proofs yield the same gradient formula.
The euclidean proof produces
formula (\ref{eq:metriceuclidean})
for the metric in terms of the euclidean two-point functions of the 
boundary fields.
The euclidean two-point functions can be written in terms of the real 
time response functions,
\eq
\expvalequil{ {\phibF(\tau)} {\phiaF(0)} }
\frac1{2\pi} \int \dif \omega \; 
\frac{\me^{-\omega\tau}}
{1+\me^{-\beta\omega}}
f_{ab}(\omega)
\en
for $0<\tau<\beta$.
Substituting in the euclidean formula (\ref{eq:metriceuclidean}) and carrying out the 
integral over $\tau$, we get the real time 
formula,
\eq
\gsusy_{ab} = 2\pi \int_{0}^{\beta}d\tau\; \sin \left ( \frac{\pi \tau}\beta  \right )
\frac1{2\pi} \int \dif \omega \; 
\frac{\me^{-\omega\tau}}{1+\me^{-\beta\omega}}f_{ab}(\omega)
=  \int d\omega\;
\frac{\pi/\beta}{\omega^{2} + \pi^{2}/\beta^{2}}
\; 
f_{ab}(\omega)
\,,
\en
so the gradient formulas are the same.

In the euclidean proof,
the choice of
the special spinor field
in the Ward identity is actually not unique.
In particular,
the real time proof can be translated into a euclidean proof
that uses
a somewhat different special spinor field than (\ref{ksp}), namely
\eqa
\hat\epsilon^{+}(x,\tau) &=&   
\left \{
\begin{array}{ll}
\hat\epsilon_{0} \cos \left [ \frac{\pi(\tau-\tau')}{\beta} \right]
& 0\le x \le \epsilon \\
\hat\epsilon_{0}
\cosh \left [ \frac{\pi (w-\epsilon-i\tau')}{\beta}\right ]
& \epsilon \le x
\end{array}
\right . \\
\hat\epsilon^{-}(x,\tau) &=&   
\left \{
\begin{array}{ll}
\hat\epsilon_{0} \cos \left [ \frac{\pi(\tau-\tau')}{\beta} \right]
& 0\le x \le \epsilon \\
\hat\epsilon_{0}
\cosh \left [ \frac{\pi (\bar w-\epsilon+i\tau')}{\beta}\right ]
& \epsilon \le x
\end{array}
\right .
\ena
This special spinor field is constant in $x$
within a collar $0\le x <\epsilon$ around the boundary,
and conformally Killing outside the collar.\footnote{Strictly
speaking, we should smooth over a small interval in 
$\epsilon>0$ so that the special spinor field becomes smooth in $x$ and 
$\tau$.  The proof is not affected by smoothing in $\epsilon$,
as can be seen, for example, in the real time equation (\ref{eq:epsdep}).}
This version of the proof perhaps has a slight advantage,
since it uses directly an explicit construction of the correlation 
functions of $\hat\theta(\tau)$ from the physical correlation 
functions of $G_{\mu r}(x,\tau)$,
by taking the limit $\epsilon\rightarrow 0$.
The dependence on canonical UV behavior is somewhat rearranged
between the two proofs,
though not in any way that seems significant.

\begin{center}
{\bf \large Acknowledgments}
\end{center}
 D.F. was supported  by the Rutgers New High Energy Theory
Center (NHETC) and by the Natural Science Institute of the University of Iceland.
 A.K. thanks the Rutgers NHETC for hospitality during the visit when most of this paper was written.

%%%%%%%%%%%%%%%%%%%%%%%%%%%%%%%%%%%%%%%%%%%%%%%%%%%%%%%%%%%%%%%%%%%%%%%%%%%%%%%%%%%%%%%%%%%%%%%%%%

\end{document}